\documentclass[twocolumn]{article}
\usepackage{times}
\usepackage{graphics}
\newcommand{\bc}{\begin{center}}
\newcommand{\ec}{\end{center}}
\newcommand{\be}{\begin{equation}}
\newcommand{\ee}{\end{equation}}
\newcommand{\ba}{\begin{array}}
\newcommand{\ea}{\end{array}}
\newcommand{\bea}{\begin{eqnarray}}
\newcommand{\eea}{\end{eqnarray}}
\newcommand{\noi}{\noindent}

\def\bitem#1\par{\noindent\hangindent1.0\parindent\hangafter=1\rm#1\par\smallskip}
\begin{document}

\twocolumn[
\phantom{.}
\vspace{-0.5cm}
\bc
{\large \bf
MODELLING OF POLARIZATION PROPERTIES \\
OF COMETARY DUST GRAINS }

\baselineskip=11.5pt
\vspace{0.5cm}
{D.A. Semenov and  N.V. Voshchinnikov}\\
Astronomy Department and Sobolev Astronomical Institute, St.~Petersburg
University, \\
St.~Petersburg, 198504 St.~Petersburg-Peterhof, Russia
\ec
]

\phantom{.}
\vspace{0.1cm}
\baselineskip=11.5pt

\bc
{\bf ABSTRACT}
\ec

   Observational data for dusty comets are summarized and systemized.
   The dependence of the linear polarization on various parameters (the
   phase angle, wavelength, etc.) is analyzed. An ensemble of aligned
   spheroidal particles having various sizes and chemical compositions
   is used to fit the linear and circular polarization
   observed for the comet Halley.

\bigskip
\bc
{\bf 1.\  INTRODUCTION}
\ec

Except for a few space missions to comets our information about cometary
dust grains comes from the analysis of their electromagnetic radiation. Like
atoms and molecules, dust particles have their own spectra. Most
spectral features are located in the infrared (IR) wavelength range.
\\

The optical properties of dust grains depend not only on
their chemical composition but also on their size distribution, shape,
the degree and direction of alignment.
This creates additional difficulties in interpretation of observations
of cometary dust.
\\

The luminosity of comets is tightly connected with the solar radiation.
Dust grains scatter or absorb it, the absorbed energy is reradiated in the
IR wavelength range. Here, we paid our attention to the scattered part of
the solar radiation.
\\

The electromagnetic waves are characterized by their intensity and polarization.
For atomic and molecular emissions the degree of linear polarization may be
theoretically predicted (Le Borgne and Crovisier, 1987). For a dust grain of
unknown shape and composition, the polarization may be additional diagnostic
tool for their investigations.
\\

Usually Mie theory for spherical particles is used for fitting the cometary
observations. This contradicts with several direct indicators of a non-sphericity
of cometary grains: for instance, non-zero degree of the linear polarization at
the phase angle $\alpha^*=0^{\circ}$ (in this case the scattering
angle $\Theta=180^{\circ}-\alpha^*=180^{\circ}$ and the Sun, the Earth and a comet are
placed on one line).
\\

In this paper, for the first time we interpret observations of the linear
polarization for the comet Halley at dif-

\phantom{.}
\vspace{0.4cm}

\noi ferent wavelengths using an ensemble of
homogeneous oriented spheroids of various sizes, chemical compositions but a
fixed shape. The degree of the circular polarization is also calculated.

\bigskip
\bc{\bf 2.\  OBSERVATIONAL DATA}\ec

\noi{\bf 2.1  Direct (in situ) measurements}
\\

The comet Halley was investigated by the Giotto and Vega 1, 2 spacecraftes.
Chemical analysis of the cometary dust has shown that the grains mainly consist
of silicates (chondrites) and carbonaceous materials. The dust experiments
discovered the presence of both large and small particles in the cometary dust.
We used the power law size distribution provided by the experiments
\be
dn \sim r^{-2.5} dr,
\ee
where $dn$ is the number of particles in the size range $[r, r+dr]$
(Mazets et al., 1986).
\\

\noi{\bf 2.2  Polarimetric observations}
\\

The observed degree of the linear polarization is determined by the two
processes: scattering of the solar radiation by cometary dust and resonance
fluorescence of molecules
\be
  P = \frac{P^{\rm sca}_{\rm dust}F^{\rm sca}_{\rm dust} +P^{\rm em}_{\rm mol}F^{\rm em}_{\rm mol}}
  {F^{\rm sca}_{\rm dust} + F^{\rm em}_{\rm mol}},
\ee
where $P^{\rm sca}_{\rm dust}$, $F^{\rm sca}_{\rm dust}$, $P^{\rm em}_{\rm mol}$, $F^{\rm em}_{\rm mol}$
are the degree of polarization and flux of the dust and gaseous components,
respectively.
In Eq.~(2)
\be
  P = P(\alpha^{*}, \lambda, D, r_{\rm H}),
\ee
where $\alpha^{*}$ is the phase angle, $\lambda$ the wavelength, $D$ the
aperture, $r_{\rm H}$ is the heliocentric distance of a comet.
\\

The phase dependence of the polarization shows features common to all comets:
it has negative branch at $\alpha^{*}<20^{\circ}$, changes the sign at phase
angle $\sim 20^{\circ}$ (the inversion angle $\alpha^{*}_{0}$)
and reaches the maximum around $90^{\circ}$ (see Fig.~1 in Jockers,
1998). The maximum degree of polarization is about $25\%$--$30\%$ for the dusty
comets and $8\%$--$15\%$ for the gaseous ones.
\\

The dusty comets are characterized by strong continuum, weak gas emissions and a
high dust-to-gas ratio. The phase curves for these comets are similiar within
the accuracy of observations. Before the appearance of the comet
Hale-Bopp, the dusty comets were believed to follow a common phase curve of
polarization. The group of the dusty comets includes the comets: 1P/Halley,
C/Hale-Bopp (1995 O1), C/Hyakutake (1996 B2), etc. The phase curve of
polarization observed for the comet Halley is typical of the comets of this type
(see Figs.~1--3).
\\

The gaseous comets have weak continuum, numerous emission lines observed
almost everywhere in the visible spectrum and a low dust-to-gas ratio.
The group of the gaseous comets includes the comets: P/Austin (1982 VI),
Kobayashi-Berger-Milon (1975 IX), Tabur (1996 Q1), etc.
\\

The degree of polarization grows with an increase of the wavelength from the
visual to IR. The main observed parameters describing the polarization of the
comet Halley are presented in Table 1.

\begin{table}[h]
\caption[]{Parameters of the phase curve for the comet Halley at different
wavelengths. Errors are about $0.5^{\circ}$ for the angles and $0.25$\% for the
polarization degree}
\begin{center}
\begin{tabular}{|l|l|l|l|l|l|}
\hline
$\lambda$ & $\alpha^{*}_{0}$ & $\alpha^{*}_{\rm min}$
& $P_{\rm min}$ & $P_{\rm max}$ & $\frac{dP}{d\alpha^{*}}\mid_{\alpha^{*}_{0}}$\\
\hline
$3650$\AA & $17^{\circ}.0$ & $?$ & $?$ & $15.5$\% & $?$ \\
\hline
$4840$ & $20.5$ & $10^{\circ}.0$ & $-1.3$\% & $16.9$ & $0.20\% / 1^{\circ}$\\
\hline
$6700$ & $23.5$ & $10.5$ & $-1.5$ & $17.1$ & $0.25\% / 1^{\circ}$\\
\hline
\end{tabular}
\end{center}
\end{table}

We used the observational data from the following papers: the degree of
polarization from Kikuchi et al. (1987), the positional angle of polarization and
circular polarization from Dollfus and Suchail (1987).
\\

Data obtained with approximately equal projected area of the aperture
centered on the nucleus of the comet were utilized. This allows us to exclude
the influence of variations of the degree and direction of polarization
observed in outer parts of coma.
\\
\begin{figure}
\bc
\resizebox{55mm}{!}{\includegraphics{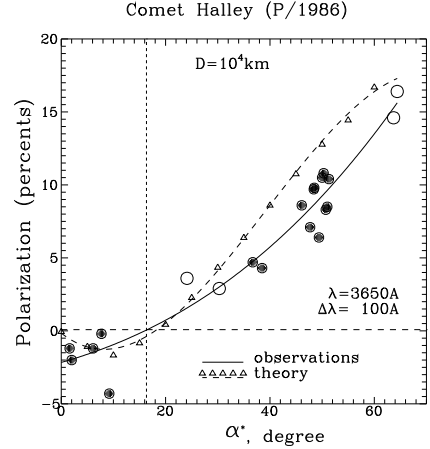}}
\caption[]{\rm Degree of the linear polarization
as a function of the phase angle $\alpha^{*}$ for
$\lambda=3650 {\rm \AA}$.
All error bars usually are less than size of signs}
\label{f1}
\ec
\end{figure}

\begin{figure}
\bc
\resizebox{55mm}{!}{\includegraphics{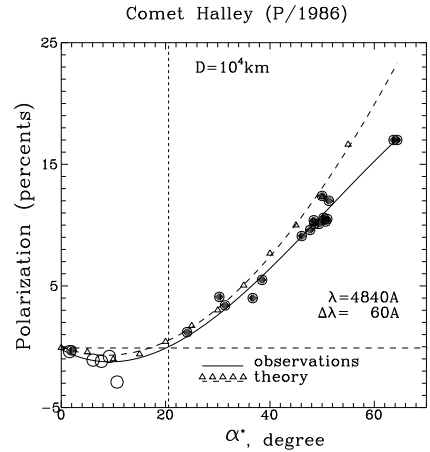}}
\caption[]{\rm The same as Fig.~1 but for $\lambda=4845 {\rm \AA}$}
\label{f2}
\ec
\end{figure}

\begin{figure}
\bc
\resizebox{55mm}{!}{\includegraphics{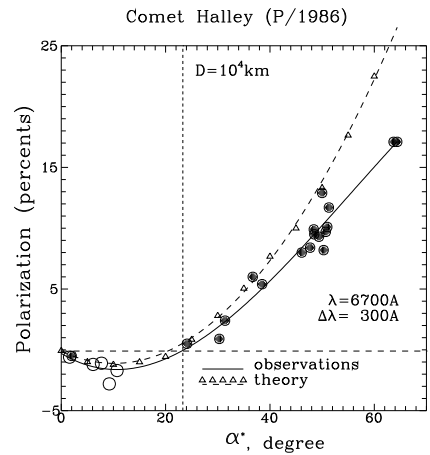}}
\caption[]{\rm The same as Fig.~1 but for $\lambda=6700 {\rm \AA}$}
\label{f3}
\ec
\end{figure}

The dependence of the degree of the linear polarization from the heliocentric
distance of a comet is masked by phase angle variations. For group of dusty
comets the degree of the linear polarization is similiar at the same phase angles
while their heliocentric distances can be different. Apparently, available
observations do not allow us to study this effect.
\\

The observations of circular polarizations of comets are very hard to do, so
they can be done only episodically. The comet Halley is an exception: the
observational data are availible for various phase angles.
\\

\noi{\bf 2.3  Non-spherical particles in comets}
\\

Single light scattering by the oriented non-spherical particles produces
phenomena which cannot be explained using the spherical particles, namely:

\begin{enumerate}
\item{the non-zero circular polarization \\
(Metz and Haefner, 1987);}

\item{rotation of the position angle with time, \\
wavelength and in different parts of a comet \\
(Dollfus and Suchail, 1987);}

\item{non-zero degree of the linear polarization at \\
the phase angle $\alpha^{*}=0^{\circ}$;}

\item{non-zero degree of the linear polarization of \\
stellar radiation observed during stellar occultations by a comet
(Rosenbush et al., 1994).}
\end{enumerate}
\begin{figure}
\bc
\resizebox{60mm}{!}{\includegraphics{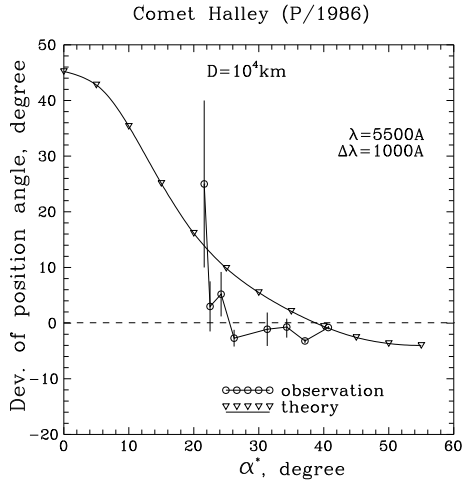}}
\caption[]{\rm Deviations of the position angle of the linear polarization
as a function of the phase angle $\alpha^{*}$}
\label{f4}
\ec
\end{figure}

\bigskip
\bc{\bf 3.\  MODEL}
\ec

For calculations we used the numerical code based on the exact solution to
the light scattering problem by the Separation of Variable Method
(Voshchinnikov and Farafonov, 1993).
The algorithm of calculations was as follows:

\begin{enumerate}
\item{calculation of an array of scattering matrices;}

\item{the averaging over rotation;}

\item{the averaging over a size distribution function;}

\item{mixing of scattering matrices for different materials;}

\item{calculations of Stokes parameters and a comparison with observations.}
\end{enumerate}

\noi{Relative errors of the calculations were usually less than 0.1\%.}
\\

The best model found has the following parameters:

$\bullet$ mixture of 80\% astronomical silicate (astrosil) \\
\phantom{........} and 20\% amorphous carbon (AC1);

$\bullet$ prolate (50\%) and oblate (50\%) spheroids;

$\bullet$ aspect ratio $a/b=2.0$;

$\bullet$ $r_{\rm V} = 0.05 - 0.25$\,$\mu$m for AC1;\footnote{
$r_{\rm V}$ is the radius of the sphere whose \
volume is equal to that of a spheroid.}

$\bullet$ $r_{\rm V} = 0.05 - 0.55$\,$\mu$m for astrosil;

$\bullet$ $\gamma=-2.5$;

$\bullet$ $\Omega=75^{\circ}$;\footnote{$\Omega$ is the angle between alignment
direction and direction of light propagation.}

$\bullet$ perfect rotational (Davies-Greenstein, DG) \\
\phantom{........} orientation.
\\

The value of $\gamma$ and the chemical composition of grains were chosen on
the basis of direct (in situ) measurements and ground-based observations. Other
parameters were estimated by fitting the observed quantaties.
\\
\begin{figure}
\bc
\resizebox{60mm}{!}{\includegraphics{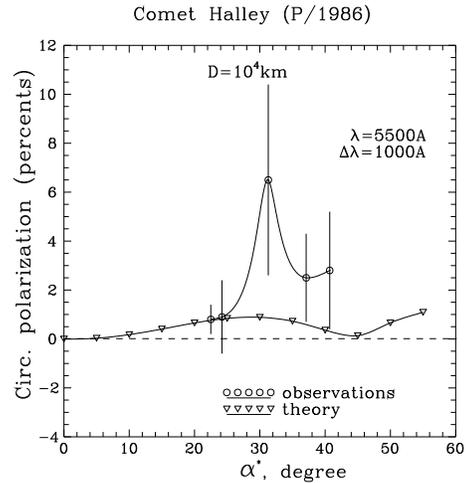}}
\caption[]{\rm Degree of the circular polarization
as a function of the phase angle $\alpha^{*}$.
All error bars are presented by vertical lines}
\label{f5}
\ec
\end{figure}
\bigskip
\bc{\bf 4.\  COMPARISON WITH OBSERVATIONS}
\ec

The observed and calculated curves are plotted in Figs.~1--5. A good
agreement with the observations is only for the negative branch of the
polarization curve, when $\alpha^{*}<30^{\circ}$ (Figs.~1--3). The degree of the
calculated polarization is higher than observed at
$\alpha^{*}>40^{\circ}-50^{\circ}$
that is connected with the high polarization efficiency of the small
amorphous carbon grains at these phase angles.
\\

Deviations of the position angle of the linear polarization plotted in Fig.~4
show that the model satisfactorily explains the observations.
The same can be said about the behaviour of the circular polarization presented
in Fig.~5. A better agreement between the theory and observations of the
circular polarization may be reached if one decreases the fraction of astrosil
in the model.

\bigskip
\bc {\bf  ACKNOWLEDGEMENTS} \ec

The authors are thankful to V.B.~Il'in for careful reading of the manuscript.
This work was financially supported by grants from the INTAS (grant 99/652)
and Volkswagen foundation.

\bigskip
\bc{\bf REFERENCES}\ec

\bitem{
Dollfus, A. and J.-L. Suchail, 1987: Polarimetry of grains in the coma of
P/Halley. I. Observations. {\it Astronomy abd Astrophysics,} {\bf 187,}
669-688.}

\bitem{
Jockers, K., 1997: Observations of scattered light from cometary dust and
their interpretation. {\it Earth, Moon and Planets,} {\bf 79,} 221-245.}

\bitem{
Kikuchi, S. et al., 1987: Polarimetry of Comet P/Halley. {\it Astronomy and
Astrophysics,} {\bf 187,} 689-692.}

\bitem{
Le Borgne, J. and J. Crovisier, 1987: Polarimetry of molecular bands in comets
P/Halley and Harley-Good.
In {\it ESA Proceedings of the 20th ESLAB symposium on the exploration
of Halley's comet,} ESA-SP-278, 571-576.}

\bitem{
Mazets, E. et al., 1987: Dust in comet P/Halley from Vega observations.
{\it Astronomy and Astrophysics,} {\bf 187,} 699-712.}

\bitem{
Metz, K. and R. Haefner, 1987: Circular polarization near the nucleus of comet
P/Halley. {\it Astronomy and Astrophysics,} {\bf 187,} 539-542}

\bitem{
Rosenbush, V. et al., 1994: Comets Okazaki-Levy-Rundenko (1989 XIX) and
Levy (1990 XX). Polarimetry and stellar occultations. {\it Icarus,} {\bf 108,}
81-91.}

\bitem{
Voshchinnikov, N. and V. Farafonov, 1993: Optical  Properties of Spheroidal
Particles. {\it Astrophysics and Space Science,} {\bf 204,} 19-86.}

\end{document}